\documentstyle[preprint,aps]{revtex}

\begin{document}


\title{Galactic Cosmic Strings as Sources of Primary Antiprotons}

\author{Glenn D. Starkman and Tanmay Vachaspati}
\address{Physics Department, Case Western Reserve University\\
Cleveland OH 44106-7079.}

\date{\today}

\maketitle

\begin{abstract}
A possible signature of a class of superconducting cosmic strings trapped
in the Milky Way plasma is the emission of low energy antiprotons due
to baryon number violating processes on the string. 
We find the terrestrial flux and the energy spectrum of such antiprotons. 
Current observational bounds on the flux of low energy antiprotons
place a {\it lower} bound on the string tension which is comparable to
that given by the electroweak scale.
\end{abstract}

\pacs{}

\narrowtext

Current particle physics models and cosmology suggest that a
number of cosmological phase transitions have occurred since the
big bang. Depending on the topology of the symmetry breakings
occurring at the phase transitions, lineal relics of the pre-phase
transition universe may have survived until the present epoch. These
``cosmic strings'' in the present universe could be detected
by their gravitational signatures \cite{avps} or by their emission
of energetic particles \cite{bhattacharjee} and 
$\gamma$-rays \cite{mohrb} if they are superheavy. 
If they are heavy and superconducting, they could be seen via 
their electromagnetic signatures \cite{witten,avgfetc}. 
In this article, we point out that the emission of antiprotons
by some\footnote{Our analysis only applies to strings on
which there are quark and lepton zero modes.}
light superconducting strings that are trapped in the galactic
plasma may yield yet another observable signature that would be
relevant for ongoing and future searches ({\it eg.} the proposed
AMS observatory) for cosmic ray antiprotons.

A number of hypothetical particle physics models containing
superconducting strings have been constructed \cite{witten,shafietal}.
String superconductivity arises because charged fermions
have zero modes on the strings which can propagate along the
string. Here we shall only be interested in superconducting
strings in which the charge carriers are ordinary quarks
and hence also carry baryon number. In this case, the
growth of electric current on the string is accompanied by the
production of baryon number on the string.
This can be seen directly by counting the number of quarks and
leptons that are produced (but still confined to the string)
when an electric field is applied along
the string \cite{witten}. The production of particles of charge
$q$ due to an electric field of magnitude $E$ along the string
is given by
\begin{equation}
{{d^2N} \over {dtdl}} = {{qE} \over {2\pi}}
\label{ntl}
\end{equation}
where, $N$ is the number of particles produced by the full
length of string. As
the charges of the electron and the $u$ and $d$ quarks are in the
ratio $-3:2:-1$, the numbers in which they
are produced are in the ratio
$-3:3\times 2: -3\times 1$. (The quarks are thrice as
numerous as the electrons because of their color degree of
freedom; the minus sign means that antiparticles are produced.)
It is easy to see that this production of particles
is also accompanied by the production of baryon number. This
follows since if 6 up quarks are produced then only 3 down antiquarks
are produced. Each quark (antiquark) has baryon number $1/3$ ($-1/3$)
and so the baryon number produced is $6/3-3/3 = 1$. If the electric
field were applied in the opposite direction, antibaryons
would be produced.

The application of an electric field along the string generates
an electric current on the string which can grow until
a critical current is reached. This critical current is given
by
\begin{equation}
i_c = \kappa (\alpha \mu )^{1/2}
\label{icrit}
\end{equation}
where $\kappa$ is a numerical constant that depends on the
details of the particle physics, $\mu$ is the mass per
unit length of the string and $\alpha$ is the fine structure
constant. The existence of a critical current follows because
when the fermions on the string occupy high enough energy
levels, they can jump off the string into the ambient vacuum.
So the critical current is roughly given by the
mass of the fermions in vacuum ($i_c \sim m_{vac}$).
If the fermions are confined, however, the correct mass to
consider would be the lowest mass
hadron. In fact, in our case, the Fermi levels can only decrease
by the leakage of baryons and so the correct mass for us to
consider is the mass of the proton $\sim 1$ GeV. With
$\mu = m_W^2 \mu_{-6} {\rm GeV}^2$ where $m_W \sim 100$ GeV
is the mass of the $W$ boson and $\mu_{-6}$ is a free
parameter, we find
$\kappa \sim 0.1 \mu_{-6}^{-1/2}$. Note that $\mu_{-6} = 1$
corresponds to strings formed at the electroweak scale with
$\mu \simeq 10^4 {\rm GeV}^2 \simeq 10^{-6} {\rm g/cm}$.

The dynamics of light superconducting cosmic strings has been
investigated in a series of papers by Vilenkin and collaborators
\cite{avgfetc,ecav,avps} leading to the following scenario.
A cosmological phase transition can lead to the formation of a network
of light, superconducting strings and the strings can be
expected to be produced with small electric currents on them. 
(If there is a primordial magnetic field present, the electric currents 
can be much stronger.) Such a network of strings will be frozen in 
the cosmological plasma in a way similar to primordial magnetic fields.
Once galaxy formation starts the strings flow into the
protogalaxy together with the ambient  plasma.
Subsequent turbulence in the galactic
flow stretches the strings and generates a tangle. 
The galactic turbulence drives the string
energy to smaller and smaller length scales (higher curvature). But,
at a critical curvature radius, $R_*$, the frozen-in 
condition breaks down and
the curved string breaks away from the ambient plasma and collapses.

There are two regimes that we need to consider depending on
if $R_*$ is larger or smaller than the
typical length scale ($L \sim 10^{20} {\rm cm}$) associated
with galactic turbulence. On scales smaller than $L$, the plasma
velocity is given by magnetohydrodynamic turbulence\cite{zelbook}:
\begin{equation}
v_R \sim v_L \left ( {R \over L} \right )^{1/4} \ , \quad (R < L)
\label{vr1}
\end{equation}
while, on scales larger than $L$ the velocity is described by the
Kolmogorov spectrum:
\begin{equation}
v_R \sim v_L \left ( {R \over L} \right )^{1/3} \ , \quad (R > L)
\label{vr2}
\end{equation}
where $v_L \sim 10^6 {\rm cm/s}$.
The string curvature scale ($R_*$) at which free collapse occurs can
be found by equating the typical string velocity 
at that scale \cite{ecav} 
\begin{equation}
v_{string} \sim {c\over \kappa R} \sqrt{{\mu \over {\alpha \rho}}}
\end{equation}
to the plasma velocity at the same scale. Here $\rho$ is the plasma
density which we take to be $10^{-25}\  {\rm g/cm^3}$. For $R_*$
smaller than $L$, this gives \cite{avgfetc}
\begin{equation}
R_* \sim 10^{16} \kappa^{-4/5} \mu_{-6}^{2/5} \ {\rm cm} \ , \quad
(R_* < L)
\label{rstar1}
\end{equation}
where, $\mu$ has been written as $10^{-6} \mu_{-6} {\rm g/cm}$.
For $R_*$ larger than $L$, we get
\begin{equation}
R_* \sim 10^{16} \kappa^{-3/4} \mu_{-6}^{3/8} \ {\rm cm} \ , \quad
(R_* > L) \ .
\label{rstar2}
\end{equation}
These calculations are valid only as long as $R_*$ is less than
the size of the galaxy, $R_{g} \sim 10$ kpc. For $\kappa \sim 1$,
this means that
our calculations are valid for strings formed at scales below
$\sim 10^{10}$ GeV. Heavier strings will not be trapped by the
galactic plasma and will have different dynamics.

The dynamics of the string during collapse is friction dominated and the
terminal velocity is given by the turbulent velocity at the
scale $R_*$\cite{ecav}:
$$
v_* \sim v_L \left ( {{R_*} \over {L}} \right ) ^{1/4} \ , \quad
(R_* < L)
$$
Plugging in numerical values, we find
\begin{equation}
v_* \sim 10^5 \kappa^{-1/5} \mu_{-6}^{1/10} \ {\rm cm/s} \ , \quad
(R_* < L)
\label{vstar1}
\end{equation}
Similarly, for the Kolmogorov case we get
\begin{equation}
v_* \sim 10^5 \kappa^{-1/4} \mu_{-6}^{1/8} \ {\rm cm/s} \ , \quad
(R_* > L) \ .
\label{vstar2}
\end{equation}
The energy of the loop during collapse
will be transformed into thermal energy of the surrounding medium.
In the very final throes of its collapse, when the size of the loop is
comparable to the thickness of the string, the loop will annihilate into
the vacuum and release its remaining energy
into various particles. (This last stage does not seem to be
of much consequence since only
a very tiny fraction of the string energy can be released this way.)

We now wish to find the baryon number released by the string network
during its evolution. While a certain section of string is frozen-in,
there is no electric field along it, and baryon number is not produced.
But once the string section becomes curved on a scale smaller than
$R_*$, it collapses under its own tension and, in doing so, cuts across
the galactic magnetic field. By Faraday's law, the traversal across
the magnetic flux is equivalent to an applied electric field along the
string and hence will generate a current and the accompanying baryon
number. The production of baryon number is independent of the
details of the collapse - it only depends on the magnetic flux,
$\Phi$, that
the string cuts across. To see this we first find the electric field
along the string:
\begin{equation}
|\vec E | = | \vec v \times \vec B | \
\label{electric}
\end{equation}
where $\vec v$ is the velocity of the string.
Inserting this expression in (\ref{ntl}) yields the total number
of particles produced during the collapse of a loop:
\begin{equation}
N = {{q} \over {2\pi  }}\int dt \oint dl |\vec v \times \vec B | 
= {{q \Phi} \over {2\pi  }} \ .
\label{nqp}
\end{equation}
The magnetic flux $\Phi$ through a loop of size $R_*$ will be estimated 
by $B_g R_*^2$ where $B_{g} = 10^{-6} B_{-6}$ G is the strength of 
the galactic magnetic field.
Therefore the baryon number produced, $Q_B$, can be estimated as
\begin{equation}
Q_B \sim \pm 0.1 \ | \Phi | \sim \pm 0.1 B_{g} R_*^2
\label{baryon0}
\end{equation}
Inserting numerical values gives\footnote{A convenient conversion
factor is $1 {\rm G-cm^2} = 3\times 10^4$.}:
\begin{equation}
Q_B \sim \pm 10^{30} \kappa^{-8/5} \mu_{-6}^{4/5} B_{-6} \ , \quad
(R_* < L )
\label{baryon1}
\end{equation}
\begin{equation}
Q_B \sim \pm 10^{30} \kappa^{-3/2} \mu_{-6}^{3/4} B_{-6} \ , \quad
(R_* > L ) \ .
\label{baryon2}
\end{equation}
The sign of the baryon number can be positive or negative depending
on the orientation of the string with respect to the galactic magnetic
field. We expect that both baryons and antibaryons will be produced
in roughly equal numbers but in different regions of the galaxy.

The time scale on which a string loop of size $R_*$ collapses is
\begin{equation}
t_* = R_* /v_* \sim  10^{11} \kappa^{-3/5} \mu_{-6}^{3/10} \ {\rm s} \ ,
\quad (R_* < L)
\label{tstar1}
\end{equation}
and,
\begin{equation}
t_* = R_* /v_* \sim 10^{11} \kappa^{-1/2} \mu_{-6}^{1/4} \ {\rm s} \ ,
\quad (R_* > L) \ .
\label{tstar2}
\end{equation}
To find the number density of antiprotons in our galaxy today
we must sum up the number of antiprotons produced
by all the loops over the entire lifetime of
the galaxy $t_g \sim 10^{17}$ s. This gives
\begin{equation}
n_{\bar p} \sim {{|Q_B |} \over {R_* ^3}} {{t_g} \over {t_*}}
\label{pbartot}
\end{equation}
The antiproton number density can be converted to a terrestrial
energy flux in antiprotons by multiplying by the velocity of the
antiprotons (order $c$). This yields:
\begin{equation}
f_{\bar p} \sim 10^2 \kappa^{7/5} \mu_{-6}^{-7/10} B_{-6} \
{\rm GeV/m^2-s}\ , \quad (R_* < L)
\label{pbarflux1}
\end{equation}
\begin{equation}
f_{\bar p} \sim 10^2 \kappa^{5/4} \mu_{-6}^{-5/8} B_{-6} \
{\rm GeV/m^2-s}\ , \quad (R_* > L) \ .
\label{pbarflux2}
\end{equation}
The dependence of these estimates on the string scale are more
transparent in the special case when only quark and lepton zero 
modes are present. Then we can use
use $\kappa \sim 0.1 \mu_{-6}^{-1/2}$  and the estimates become
\begin{equation}
f_{\bar p} \sim 10 \mu_{-6}^{-7/5} B_{-6} \
{\rm GeV/m^2-s}\ , \quad (R_* < L)
\label{flux1}
\end{equation}
\begin{equation}
f_{\bar p} \sim 10 \mu_{-6}^{-5/4} B_{-6} \
{\rm GeV/m^2-s}\ , \quad (R_* > L) \ .
\label{flux2}
\end{equation}
Note that
the antiproton flux {\it decreases} with increasing string
tension. So heavier
strings produce fewer antiprotons and the largest flux of
antiprotons is due to the lightest strings. Assuming that
the standard model is correct up to the electroweak scale
we have 
$\mu_{-6} > 1$ and this gives us an upper bound on the
antiproton flux from strings.

These estimates assume that the antiprotons do not leak out
of the galaxy volume
and neither are they annihilated in scatterings off protons. 
Both assumptions can be justified by rough estimates. 
The leakage timescale,
$\tau_{leak}$,
is found as the time taken by a relativistic antiproton (speed $\sim c$)
to cover a distance equal to the size of the galaxy $L_g \sim 10$ kpc.
A crucial factor that needs to be accounted for is that the antiproton
trajectory is not a straight line but piecewise circular due to the galactic
magnetic field. Assuming a random walk for the antiproton with step
size similar to the cyclotron radius 
$
\xi \sim {{mc}/ B_g}
$ 
we find 
$$
\tau_{leak} \sim {{L_{g}^2} \over \xi} \sim 10^{19} \  {\rm s}
$$ 
which is longer than the age of the galaxy.
The scattering
timescale, $\tau_{scatt}$, of the antiprotons is found by using the
proton-antiproton scattering crosssection $\sigma_{p {\bar p}} \sim
m_p ^{-2}$ where $m_p$ is the mass of the proton
and the galactic proton density $\sim 10^{-24} {\rm g/cm^3}$.
This gives 
$$
\tau_{scatt} \sim {{m_p} \over {\rho \sigma_{p {\bar p}}} } 
\sim 10^{18} \ {\rm s}
$$
and so the survival time for the antiprotons is longer than the
age of the galaxy.

So far we have only considered the total flux of antiprotons.
We now turn to the antiproton energy distribution.
The antiprotons that come off the string due to $u$
and $d$ quarks on the string will have energy of order the
proton mass. We might think that
the $c$ and $s$ quarks will come off
at energies when they can combine into baryons and that the $t$
and $b$ quarks can only jump off the string at energies greater than
the top quark mass ($\sim 200$ GeV). If true, the string would emit 
a third of the baryons at energies of a few hundred
GeV which would then decay into protons and antiprotons
of similar energies. This would be quite a distinctive signature for
strings. However, this expectation ignores flavor changing processes
by which $t$ quarks living on the string, scattering off $d$ quarks
living on the string,  can be emitted as $d$ quarks and $u$ quarks
living off the string. The relevant terms in the Lagrangian for such an 
interaction are
$$
{g \over 2} U_{td} {\bar d}_L^{(1)} \gamma^\mu t_L^{(0)} W_\mu^- \ ,
\quad
{g \over 2} U^\star_{du} {\bar u}_L^{(1)} \gamma^\mu d_L^{(0)} W_\mu^+
$$
where the superscripts refer to whether the fermionic field is
a zero mode (0) or a massive mode (1) and the quantity $U_{ij}$
is the $ij$ component of the CKM matrix (see 
\cite{chengli} for example). So the interaction rate is suppressed
due to the factor $\vert U_{td}\vert^2 \sim 10^{-4}$. In addition, there is a
geometric suppression since only a fraction of the $d_L^{(1)}$ wavefunction
overlaps with the ${\bar t}_L^{(0)}$ zero-mode wavefunction. This amounts
to a suppression by $(m_d / m_t)^2$ in the cross-section; a less important
final-state wave-function overlap is estimated as $(m_u/m_d)^2$. 
Combining
these factors with the Fermi cross-section, and a quark number density  
on the string, $n_q \simeq p_F m_q^2,$ gives a time scale for
flavor changing interactions:
\begin{equation}
\tau_{flavor} \sim \left[{G_F^2 E^2\over 4 \pi^2} 
\vert U_{td}\vert^2 \vert U_{du}\vert^2 
\left ( {{m_d} \over {m_t}} \right )^2
\left ( {{m_u} \over {m_d}} \right )^2
(p_F m_d^2)\right]^{-1}
 \sim 10^{3-4} \ {\rm s} 
\label{scatt}
\end{equation}
Actually, of course, the light quarks come off the string not as bare
quarks, but as pions, with the baryon number remaining on the string.
Thus, for example, $t^{(0)} + d^{(0)} \longrightarrow 
u^{(0)} + d^{(0)} + 2\pi^0$.
We have neglected the QCD corrections to the final state.

This flavor changing rate is somewhat faster than
the astrophysical rate associated with the build up of ($1$ GeV of) Fermi
momentum on the string as the string cuts through the galactic magnetic 
field:
$$
\tau_{Fermi} \sim {1 GeV \over {\dot p}_F} 
\sim {1GeV \over q B v}  
\sim {10^{11}{\rm cm}\over q B_{-6} v} 
$$
Using (\ref{vstar1}) and (\ref{vstar2}) this gives,
$$
\tau_{Fermi} \sim 10^6 \kappa^{1/5} \mu_{-6}^{-1/10} {\rm s}\ , 
\quad (R_* < L)
$$
$$
\tau_{Fermi} \sim 10^6 \kappa^{1/4} \mu_{-6}^{-1/8} {\rm s}\ , 
\quad (R_* > L)
$$
This implies that the $t$ quarks will convert to $d$ quarks and escape
from the string as $d$ quarks 
and the threshold for this to happen is $\sim 1$ GeV
(instead of $\sim 174$ GeV). Therefore, the Fermi momentum on the string
will saturate at about 1 GeV and only mildly relativistic protons and 
antiprotons will be emitted from the string\footnote{For large
$\mu_{-6}$ it may happen that $\tau_{Fermi} < \tau_{flavor}$ but
then the total flux of antiprotons, as given by eqns. (\ref{flux1})
and (\ref{flux2}), is too small to be of interest.}.

Once the antiprotons are ejected from the string, we expect them to
undergo acceleration by the usual mechanism of shock acceleration
proposed by Fermi \cite{fermi,gaisserbook}. 
This would lead to a spectrum
of antiprotons whose energy distribution falls off as $E^{-2}$ and
which is normalized by the total flux given in (\ref{flux1}) and
(\ref{flux2}). If the antiproton energy flux at energy $E$ is 
denoted by $F(E)$, we have
\begin{equation}
dF = f_{\bar p} {{m_p} \over E} {{dE} \over E}
\label{F1}
\end{equation}
where $f_{\bar p}$ is given in eqns. (\ref{pbarflux1}) 
and (\ref{pbarflux2}) and
$m_p$ is the mass of the proton.

The currently favored explanation for the observed antiprotons
in cosmic rays is that they are produced as secondary particles 
due to cosmic ray collisions off interstellar matter and
present observations are consistent with this scenario 
(see \cite{gaissershaefer} for a review and \cite{yoshimura} 
for recent observations of low energy antiprotons). 
If we assume that the antiprotons seen so far are all
secondary particles, then the current observations place an 
upper bound on the primary flux. 
The present bound for the antiproton flux in the kinetic 
energy range between 1 to 4 GeV is $dF < 10 {\rm GeV/m^2-s}$. 
So the antiproton flux from galactic superconducting strings 
produced at scales larger
than the electroweak scale is not constrained by present
observations.  But continued observations and the 
proposed AntiMatter Search Observatory will either produce 
positive evidence for this possible exotic 
constituent of our Milky Way or impose interesting constraints. 

\acknowledgments

We are grateful to Tom Gaisser and Dick Mewaldt for advice and comments.

%
%
%
%
%


\begin{references}

\bibitem{avps} A. Vilenkin and E.P.S. Shellard,
{\it Cosmic Strings and other Topological Defects}
(Cambridge University  Press, Cambridge, 1994).

\bibitem{bhattacharjee} P. Bhattacharjee, C.T. Hill and D.N. Schramm,
Phys. Rev. Lett. {\bf 69}, 567 (1992).

\bibitem{mohrb} M. Mohazzab and R. Brandenberger, Int. J. Mod.
Phys. {\bf D2}, 183 (1993).

\bibitem{witten} E. Witten, Nucl. Phys. {\bf B249}, 557 (1985).

\bibitem{avgfetc} E.M. Chudnovsky, G.B. Field, D.N. Spergel and
A. Vilenkin, Phys. Rev. {\bf D34}, 944 (1986).

\bibitem{shafietal} G. Lazarides, C. Panagiotakopoulos and
Q. Shafi, Phys. Rev. Lett. {\bf 56}, 432 (1987); Phys. Lett.
{\bf B183}, 289 (1987).

\bibitem{ecav} E. Chudnovsky and A. Vilenkin, Phys. Rev. 
Lett. {\bf 61}, 1043 (1988).

\bibitem{chengli} T. P. Cheng and L. F. Li, {\it Gauge Theory of
Elementary Particle Physics} (Oxford University Press, 1991).

\bibitem{fermi} E. Fermi, Phys. Rev. {\bf 75}, 1169 (1949).

\bibitem{gaisserbook} T.K. Gaisser, {\it Cosmic Rays and
Particle Physics} (Cambridge University Press, Cambridge, 1990).

\bibitem{gaissershaefer} T.K. Gaisser and R.K. Schaefer,
Ap. J. {\bf 394}, 174 (1992).

\bibitem{zelbook} Ya.B. Zel'dovich, A.A. Ruzmaikin and 
D.D. Sokoloff, {\it Magnetic Fields in Astrophysics} (Gordon
and Breach, New York, 1983).

\bibitem{yoshimura} K. Yoshimura {\it et al.}, Phys. Rev. Lett.
{\bf 75}, 3792 (1995).

\end{references}
\end{document}